\input phyzzx
%\input epsf.def
%\epsfverbosetrue
\hfuzz 50pt

%%%%%%%
\font\mybb=msbm10 at 12pt
\def\bb#1{\hbox{\mybb#1}}

\def\bR {\bb{R}}
\def\bE {\bb{E}}

\def\bfomeg{\omega\kern-7.0pt\omega}
\def\bfOmeg{\Omega\kern-8.0pt\Omega}

\def\os{{\buildrel  o\over s}}
\def\of{{\buildrel  o\over f}}

%%%%%%%%%%%%

%%%%%%%%%%%%%%%%%%%%%%%%%%%%%%%%%%%%%%%%%%%%%%%%%%%%%%%%%%%%%%%%%%%%%%%%%%%%%
\REF\witten{E.Witten {\sl Solutions of four-dimensional 
field theories via M-theory}, Nucl. Phys.
{\bf B500} (1997) 3.}
\REF\ptg{J. P. Gauntlett, G.W. Gibbons, 
G. Papadopoulos \& P.K. Townsend, {\sl
Hyper-K\"ahler Manifolds and Multiply 
Intersecting Branes}, Nucl. Phys. {\bf B500} (1997)
133.} 
\REF\lib{R. R. Khuri, {\sl A comment on String Solitons},
 hep-th/9305143.}
\REF\jerome{J.P. Gauntlett, D.A. Kastor \& J. 
Traschen, {\sl Overlapping Branes in M-theory}
{\sl Nucl. Phys. } {\bf B478} (1996) 544.}
\REF\gppkt{G. Papadopoulos \& P.K. Townsend,  
{\sl Intersecting M-branes}, Phys. Lett.
 {\bf B380} (1996) 273.}
\REF\tseytlinb{A.A. Tseytlin, {\sl Harmonic 
Superposition of M-Branes}, Nucl. Phys.{\bf B475}
(1996) 149.}
\REF\gpgpb{G. Papadopoulos \& P.K. Townsend, 
{\sl Kaluza-Klein on the brane}, Phys. Lett.
{\bf B393} (1997) 59.}
\REF\leigh{M. Berkooz, M.R. Douglas \& R.G. Leigh, 
{\it Branes Intersecting at
Angles}, Nucl. Phys.  {\bf B480} (1996) 265.}
\REF\cvetic { K. Behrndt \& M. Cveti\v c, 
{\sl BPS-Saturated Bound States of Tilted
p-Branes in Type II String Theory}, Phys. Rev. 
{\bf D56} (1997) 1188.}
\REF\myers{J.C. Breckenridge, G. Michaud and
 R.C. Myers, {\sl New Angles on Branes}, Phys. Rev. 
{\bf D56} (1997) 5172;
hep-th/9703041.}
\REF\costa{ M. Costa \& M. Cveti\v c, 
{\sl Non-Threshold D-Brane Bound States and
Black Holes with Non-Zero Entropy},  Phys. Rev. {\bf D56} 
(1997) 4834; hep-th/9703204.}
\REF\larsen{V. Balasubramanian, F. Larsen \& R.G. Leigh, 
{\sl Branes at Angles and
Black Holes}, Phys. Rev. {\bf D57} (1998) 3509;  hep-th/9704143.}
\REF\myersb{G. Michaud and R.C. Myers, {\sl Hermitian 
D-Brane Solutions}, Phys. Rev. {\bf D56} (1997) 3698;
hep-th/9705079.}
\REF\gpt{G. Papadopoulos \& A. Teschendorff, 
{\it Instantons at Angles},
Phys. Lett. {\bf B419} (1998) 115; hep-th/9708116.}
\REF\shenker{T. Banks, W. Fischer, S.H. Shenker and L. 
Susskind, {\sl M Theory as
a Matrix Model}, Phys. Rev. {\bf D55} (1997) 5112; hep-th/9610043.}
\REF\zhou{ N. Ohta and J-G. Zhou, {\sl Realization of D4-Branes at Angles
in Super Yang-Mills Theory}, {\bf B418} (1998) 70; hep-th/9709065.}
\REF\kent{E. Corrigan, P. Goddard \& A. Kent, {\sl Some Comments 
on the ADHM Construction
in 4k Dimensions}, Commun. Math. Phys. {\bf 100} (1985) 1.}
\REF\ohtatown{N. Ohta \& P.K. Townsend, {\sl Supersymmetry of M-branes
at Angles}, Phys. Lett. {\bf B418} (1998) 77; hep-th/9710129.}
\REF\callan{C. G. Callan,Jr, J.A. Harvey, A. Strominger,
{\sl Supersymmetric String Solitons}, hep-th/9112030; {\sl World
Sheet Approach to Heterotic Instantons and Solitons}, Nucl. Phys. 
{\bf B359} (1991) 611.}
\REF\hull{S.J. Gates, C.M. Hull \& M. Ro\v cek, 
{\sl Twisted Multiplets and New
Supersymmetric Nonlinear Sigma Models}, Nucl. Phys. 
{\bf B248} (1984) 157.}
\REF\howea {P.S. Howe \& G. Papadopoulos, {\sl Ultraviolet
 Behaviour of Two-Dimensional
Nonlinear Sigma Models}, Nucl. Phys. {\bf B289} (1987) 264; 
{\sl Further Remarks on the
Geometry of Two-Dimensional Nonlinear Sigma Models}, Class. 
Quantum Grav. {\bf 5} (1988)
1647.}
\REF\howeb{P.S. Howe \& G. Papadopoulos, {\sl Finiteness 
and Anomalies in (4,0)
Supersymmetric Sigma Models},  Nucl. Phys. 
{\bf B381} (1992) 360. }
\REF\howec{P.S. Howe \& G. Papadopoulos, 
{ \sl Twistor Spaces for HKT manifolds},
Phys. Lett.  {\bf B379} (1996) 80.}
\REF\strominger{A. Strominger, {\sl Superstrings with Torsion}, Nucl.
Phys. {\bf B274} (1986) 253.}
\REF\gpat{G. Papadopoulos \& A. Teschendorff, manuscript in preparation.}
\REF\marica{ S.W. Hawking \& M. Taylor-Robinson, {\sl Bulk 
Charges in Eleven Dimensions}, hep-th/9711042.}
\REF\lambert{N.D. Lambert, {\sl Heterotic p-branes from
 massive sigma models }, Nucl. Phys.
{\bf B477} (1996) 141.}

%%%%%%%%%%%%%%%%%%%%%%%%%%%%%%%%%

%%%%%%%%%%%%%%%%%%%%%%%%%%%%%%%%%%%%%%%%%%%%%%%%%%%%%%%%%%%%%%%%%%%%

\Pubnum{ \vbox{ \hbox{DAMTP-R/98/22} } }
\date{June 1998}
\pubtype{}

\titlepage

\title {\bf Multi-angle Five-Brane Intersections}

\author{G. PAPADOPOULOS and A. TESCHENDORFF}
\address{DAMTP,\break University of Cambridge,
\break
Silver Street,
\break
Cambridge CB3 9EW, U.K.}
%\andauthor{}
%\address{}

\abstract{We
 find new solutions of IIA supergravity
 which have  the interpretation of
intersecting NS-5-branes at $Sp(2)$-angles
 on a string preserving at least $3/32$ of supersymmetry.  
We show that the relative
position of every pair  of NS-5-branes
involved in the superposition is determined
by four angles. In addition
 we  explore the related configurations
in IIB strings and M-theory. }

\endpage
\pagenumber=2
%\sequentialequations

\chapter{Introduction}

Most of the recent progress in understanding
 the various dualities of superstring
theories as well as their applications in
 black holes and superymmetric Yang-Mills
is due to the investigation of intersecting 
brane configurations. There are many ways to
view such configurations.  One way is 
 as classical solutions of supergravity theories
which are the effective theories of
 superstrings and M-theory. There are many
such configurations. Here we shall 
be mainly concerned with the M-theory configuration
which has the interpretation of 
M-5-branes intersecting on a string. This configuration
is related via reduction  and via 
IIA/IIB T-duality to a large number
of ten-dimensional configurations. 
These  include those of [\witten]
that have been used to investigate  
four-dimensional N=2 supersymmetric Yang-Mills
 as well as configurations that have many 
novel properties like those
of [\ptg] (and refs within).

There are many ways to superpose  two M-5-branes so as to 
intersect on a string preserving a
proportion of  spacetime supersymmetry. The
simplest case is that of  
orthogonally intersecting M-5-branes, i.e.
$$
\eqalign{
{\rm (i)\, M-5:}&\qquad 0,\,1,\,2,\,3,\,4,\,5,\,*,\,*,\,*,\,*,\,*
\cr
{\rm (ii)\, M-5:}&\qquad 0,\,1,\,*,\,*,\,*,\,*,\,6,\,7,\,8,\,9,\,*
\ .}
\eqn\inttwo
$$
In this notation, the M-5-branes are in the directions 
$0,1,2,3,4,5$ and $0,1,6,7,8,9$,
respectively, and the string is in the directions
$0,1$. The associated 
supergravity solution was found in [\lib] and its 
interpretation within M-theory was
given in [\jerome]. This solution depends on two 
harmonic functions. However, unlike  other
intersecting M-brane configurations [\gppkt, \tseytlinb, \gpgpb] the harmonic 
function associated with one of the
M-5-branes depends on the worldvolume coordinates of 
the other brane which are transverse to the
common intersection. The rest of the configurations
 involve superpositions
of M-5-branes at angles [\leigh]. For this, one of the 
M-5-branes is rotated relative
to the other with an element of a subgroup $G$ of $SO(8)$.    
An example of a supergravity solution with the 
interpretation of M-5-branes
intersecting at $Sp(2)$-angles on a string was
 given in [\ptg].  This was achieved
by T-dualizing twice the product of an 
eight-dimensional  toric hyper-K\"ahler
 metric with
a two-dimensional Minkowski one
 (as solution of IIA supergravity) and 
then lifting the
resulting solution to eleven 
dimensions\foot{For other brane intersections
at angles see [\cvetic-\myersb].}. The M-theory
configuration found in [\ptg] has the properties that
there is 
{\sl one independent angle}
between every pair of intersecting M-5-branes and
it preserves $3/16$ of  supersymmetry.

The intersecting M-5-brane configuration in [\ptg] 
is a special case of a 
more general solution for which there
are {\it four independent angles} between every pair 
of M-5-branes. One indication that more general solutions
exist from those of [\ptg] was given in [\gpt] by comparing 
Yang-Mills configurations with
M-5-brane configurations in
 the context of matrix theory [\shenker]. 
In matrix theory, longitudinal
M-5-branes correspond to four-dimensional Yang-Mills 
instantons. Using this, it turns
out that  intersecting M-5-branes at
$Sp(2)$-angles on a string  correspond to 
 Yang-Mills configurations for which the
curvature two-form\foot{The gauge group 
is $U(N)$ for some $N$.} is in the
$sp(2)$ subalgebra of $so(8)$ [\zhou, \gpt]. Interpreting a
class of solutions of this Yang-Mills BPS [\kent] 
condition as a superposition
of four-dimensional instantons, it
 was found in [\gpt] that there are {\it four
independent angles} between every 
pair of four-dimensional instantons.
Some more indications for the existence of {\sl other} supersymmetric
 multi-angle intersecting
brane solutions were also given in [\ohtatown]. This 
was done by investigating
the supersymmetry projections associated with two M-5-branes.

In this paper we shall present a method
 to superpose IIA NS-5-brane
solutions that yields new solutions of
 IIA supergravity with the
interpretation of intersecting IIA 
NS-5-branes at $Sp(2)$-angles on a string.  They generalize those of [\ptg].
We shall find that there are four independent angles between {\sl every}
pair of NS-5-branes involved in the intersection and that our solutions
preserve at least $3/32$ of supersymmetry. We shall
also show that for a {\sl given} pair of NS-5-branes 
and generic asymptotic metric, the
angles matrix has two independent eigenvalues.
We shall
then present the superpositions 
of these solutions with a fundamental string
and a pp-wave. We shall
investigate the IIB duals of these 
configurations as well as their M-theory
interpretation.

\chapter{ IIA NS-5-branes at angles}

The reduction of intersecting M-5-branes 
 on a string configuration 
of eleven-dimensional supergravity in a 
direction transverse to the branes leads to  a IIA string 
configuration with the
 interpretation of intersecting
NS-5-branes on a string. 
It turns out that it is more convenient to carry out our
computations in the context of IIA supergravity. The 
M-theory configurations can then be found by simply
lifting the ten-dimensional ones to eleven dimensions.

We are interested in a solution of IIA supergravity that
involves only NS-5-branes. This allows us to set all
Ramond$\otimes$Ramond fields
 to zero since this is a consistent
truncation of IIA supergravity. The resulting 
field equations in the string frame are
$$
\eqalign{
R_{MN}-H_{MPQ} H_N{}^{PQ}+2 \nabla_M\partial_N \phi&=0
\cr
\nabla_P \big(e^{-2\phi} H^{PMN}\big)&=0\ ,}
\eqn\fieldeq
$$
where $\phi$ is the dilaton and $H$ is the 3-form field
strength of the NS$\otimes$NS  sector, $\nabla$ is the Levi-Civita
connection of the metric $g$ and  $M,N, P,
Q=0,\dots 9$. We raise and lower indices with the metric $g_{MN}$.
There is also a third field 
equation associated with the dilaton.
However it is well known that this field equation is implied from
those in \fieldeq\ up to a constant.

The IIA NS-5-brane solution [\callan] is
$$
\eqalign{
ds^2&=ds^2(\bE^{(1,5)})+ F ds^2(\bE^4)
\cr
H&=-{1\over2} \star dF
\cr
e^{2\phi}&=F}
\eqn\fivebsol
$$
where 
$$
F=1+\sum_i{\mu_i\over |y-y_i|^2}
\eqn\mmtwo
$$ 
is a harmonic function on $\bE^4$ and the Hodge star is
that of
$\bE^4$.

We are seeking  solutions of  IIA supergravity
with the interpretation of intersecting 
NS-5-branes at angles on a string. Motivated by
\fivebsol, we  write the ansatz
for the metric and three-form field strength as
$$
\eqalign{
ds^2&=ds^2(\bE^{(1,1)})+ds^2_{(8)}
\cr
H&=H_{(8)}
\cr
e^{8\phi}&=g_{(8)}\ ,}
\eqn\ans  
$$
where $ds^2_{(8)}$ is an eight-dimensional
 metric,   $H_{(8)}$ is a closed three-form that depends on 
the same coordinates as those
of $ds^2_{(8)}$ and $g_{(8)}$ is the determinant of $ds^2_{(8)}$.
The coordinates in
$\bE^{(1,1)}$ are the worldvolume directions 
of the string lying at the
intersection\foot{We remark that there is no 
worldvolume soliton on the NS-5-brane
associated with this intersection.}. 

The main task is to determine $ds^2_{(8)}$ and $H_{(8)}$. For 
this we consider the linear
maps\foot{These linear maps are chosen such that 
they preserve certain quaternionic 
structures on $\bE^4$ and $\bE^8$. },
$\tau$, from
$\bE^8$ into $\bE^4$ given by
$$
y^\mu=p_{i\lambda} x^{i\lambda} \delta^\mu{}_0+ 
\delta^\mu{}_a (J^a){}^\lambda{}_\rho
p_{i\lambda} x^{i\rho}- a^\mu
\eqn\mmmfive
$$
where $\{y^\mu; \mu=0,1,2,3\}$ are the standard 
 coordinates of $\bE^4$,
$\{x^{i\mu}; i=1,2, \mu=0,1,2,3\}$ are the 
standard  coordinates\foot{The Euclidean
metric on $\bE^8$ is $ds^2=\delta_{\mu\nu}
\delta_{ij} dx^{i\mu} dx^{j\nu}$.} of
$\bE^8$, and
$\{J^a; a=1,2,3\}$ are constant complex structures on $\bE^4$ 
associated with a basis
of anti-self-dual two-forms. The real numbers 
$\{p_{i\mu}, a^\mu; \mu=0, \dots, 3, i=1,2\}$ are the parameters
of the linear maps. The  parameters $\{p_{i\mu}\}$, eight in total, 
are rotational while
the parameters $\{a^\mu; \mu=0,\dots,3\}$, four in total,  are 
translational.  Next using the
above linear map, we pull back
the metric 
$$
d\os^2= {1\over |y|^2} \delta_{\mu\nu}dy^\mu dy^\nu
\eqn\typhkta
$$
and the closed three-form
$$
(\of_3)_{\mu\nu\rho}=-{1\over2}\epsilon_{\mu\nu\rho}{}^\lambda 
\partial_\lambda{1\over |y|^2}
\eqn\typhktb
$$
from $\bE^4-\{0\}$ to $\bE^8$, where 
$\epsilon$ is the volume form of $\bE^4$ 
with respect to the flat metric and
$\epsilon_{\mu\nu\rho}{}^\lambda=
\epsilon_{\mu\nu\rho\sigma}\delta^{\sigma\lambda}$ and 
the norm $I\cdot|$ is with
respect to the Euclidean metric on $\bE^4$. We remark that
 \typhkta\ and \typhktb\ is the NS-5-brane geometry above after
 removing the identity in the harmonic
function $F$.  The
pulled back metric is
$$
ds^2={A_{ij}\delta_{\mu\nu}+B^s_{ij} I_{s\mu\nu}\over |p_{i\lambda} 
x^{i\lambda} \delta^\sigma{}_0+ \delta^\sigma{}_a J^a{}^\lambda{}_\rho
p_{i\lambda} x^{i\rho}- a^\sigma|^2} dx^{i\mu} dx^{j\nu}\ ,
\eqn\mmsix
$$
where $\{I_s; s=1,2,3\}$ are the constant
 complex structures on $\bE^4$ associated
with a basis of self-dual two-forms\foot{We remark that
$I_{r\mu\nu}=\delta_{\mu\lambda} 
I_r{}^\lambda{}_\nu$.} and
$$
\eqalign{
A_{ij}&=p_i\cdot p_j=\delta^{\mu\nu} 
p_{i\mu} p_{j\nu}
\cr
B^s_{ij} I_{s \mu\nu}&=p_{i\mu} p_{j\nu}
-p_{i\nu} p_{j\mu}+\epsilon_{\mu\nu}{}^{\rho
\sigma} p_{i\rho} p_{j\sigma}\ .}
\eqn\mmseven
$$
To derive the above expressions, we have used the identity
$$
 J^r{}_{\mu\nu} J_r{}_{\alpha\beta}=\delta_{\mu\alpha}
\delta_{\nu\beta}-\delta_{\nu\alpha} 
\delta_{\mu\beta}-\epsilon_{\mu\nu\alpha\beta}\ .
\eqn\mmeight
$$

The metric and closed three-form in the ansatz
\ans\  are constructed by summing the 
pull-backs of $d\os^2$ and $\of_3$ over different
choices of linear maps
$\tau$, i.e.
$$
\eqalign{
ds^2_{(8)}&=ds^2_{\infty}+\sum_\tau \mu(\tau) \tau^* d\os^2
\cr
H_{(8)}&=\sum_\tau \mu(\tau) \tau^* \of_3\ ,}
\eqn\fhkt
$$
where 
$$
ds^2_{\infty}=\big(U^\infty_{ij} \delta_{\mu\nu}+ 
(V^\infty)^r_{ij} I_{r \mu\nu}\big)
dx^{i\mu} dx^{j\nu}
\eqn\mmnine
$$ 
is a constant metric on $\bE^8$ and $\mu(\tau)$ are real
 positive numbers. Observe that
 $ds^2_{(8)}\rightarrow ds^2_{\infty}$  as
$|x^i|\rightarrow
\infty$, i.e. $ds^2_{\infty}$ is the 
asymptotic metric of $ds^2_{(8)}$.   
Explicitly, the metric $ds^2_{(8)}$ is
$$
\eqalign{
ds^2_{(8)}&\equiv g_{i\mu, j\nu} 
dx^{i\mu} dx^{j\nu}\equiv \big(U_{ij}\delta_{\mu\nu}+
V^s_{ij} I_{s\mu\nu}\big) dx^{i\mu} dx^{j\nu} 
\cr &
=ds^2_{\infty}+\sum_{\{p, a\}} \mu(\{p,a\})
{A(p)_{ij}\delta_{\mu\nu}+B(p)^s_{ij} I_{s\mu\nu}\over
|p_{i\lambda}  x^{i\lambda} \delta^\sigma{}_0
+ \delta^\sigma{}_a J^a{}^\lambda{}_\rho
p_{i\lambda} x^{i\rho}- a^\sigma|^2} dx^{i\mu} dx^{j\nu}\ .
}
\eqn\mmten
$$
The  expression of  $H=H_{(8)}$ and  dilaton $\phi$ in terms of $U, V_r$ is
$$
\eqalign{
H_{i\mu, j\nu, k\rho}= {1\over 3!} \big [ -3 \epsilon_{\mu\nu\rho}{}^\tau &
\partial_{(i\tau} U_{jk)}
+ 2 \big\{\delta_{\mu\nu} (\partial_{[i\rho} U_{j]k}+  I_s{}^\tau{}_{ \rho}
\partial_{[i\tau} V^s_{j]k}) 
\cr &
- I^s_{\mu\nu} \partial_{(i\rho} V^s_{j)k}+ {\rm cyclic }\ \
(i\mu,j\nu,k\rho)\big\} \big] }
\eqn\torsionbig
$$
and
$$
e^{8\phi}=\big({\rm det}(U_{ij})-
\sum^3_{r=1}{\rm det}(V^r_{ij})\big)^4\ ,
\eqn\dilbigg
$$
respectively. We have  verified using 
$$
g^{i\mu, j\nu}\partial_{i\mu} g_{j\nu,k\lambda}={1\over4} g^{i\mu, j\nu}
\partial_{k\lambda} g_{i\mu, j\nu}
\eqn\keyrel
$$
and some of the results in [\strominger] that the above configuration
is a solution of the IIA supergravity field equations.

Our solutions preserve at least $3/32$ of supersymmetry. To see this,
we introduce the complex structures 
$$
{\bf J}_r{}^{i\mu}{}_{j\nu}= J_r{}^\mu{}_\nu \delta^i{}_j
\eqn\mnone
$$
on $\bE^8$ and $r=1,2,3$. Then after some computation
one can  show that these complex structures are covariantly constant
with respect to the connection
$$
\nabla^{(+)}=\nabla + H\ ,
$$
where $\nabla$ is the Levi-Civita connection of $ds^2_{(8)}$.
This implies that for generic choices of linear maps $\tau$ the holonomy of
 $\nabla^{(+)}$ is exactly\foot{Therefore, the
eight-dimensional geometry $(ds^2_{(8)}, H_{(8)})$
 admits an hyper-K\"ahler
structure with torsion [\hull-\howec]  with respect to the pair 
$(\nabla^{(+)}, {\bf J}_r)$.} 
$Sp(2)$. Using this, one can show that the above
 solution admits at least three Killing
spinors and therefore it preserves at least
$3/32$ of the supersymmetry\foot{More details 
will be given elsewhere [\gpat].}.

Our solution includes that of [\ptg]. To 
find the latter, 
we repeat the construction as
 before but in this case we sum over
linear maps $\tau$ with rotational parameters
$$
\{p_{i\mu}\}=\{p_i, 0,0,0\}\ .
\eqn\mnfive
$$
It is then straightforward to show that 
the solution \ans\ reduces to that of
[\ptg].

\chapter{Angles}

The new solution 
\ans\ of IIA supergravity has been
constructed as a superposition of 
NS-5-branes intersecting on a string.
To find the
angles amongst any two NS-5-branes involved in 
the superposition, we take the location
of the branes involved in the configuration
 to be determined by the poles
of the metric \fhkt, i.e. the kernels of the linear
 maps $\tau$. Given two such
maps depending on the parameters $(\{p_i\}, a)$ 
and $(\{q_i\}, b)$, respectively, the
kernels are given by the equations
$$
\eqalign{
p_{i\lambda} x^{i\lambda} \delta^\mu{}_0+
 \delta^\mu{}_a J^a{}^\lambda{}_\rho
p_{i\lambda} x^{i\rho}- a^\mu&=0
\cr
q_{i\lambda} x^{i\lambda} \delta^\mu{}_0+ 
\delta^\mu{}_a J^a{}^\lambda{}_\rho
q_{i\lambda} x^{i\rho}- b^\mu&=0\ .}
\eqn\poles 
$$
It is straighforward to observe that each of the four-dimensional planes
associated with the above equation depend on at most four rotational
parameters up to a redefinition of the translational ones\foot{This is 
also true for the
metric and three-form field strenth of the solution. It turns out 
that they depend on at
most four rotational parameters for every five-brane
 involved in the intersection.}. The
normal vectors of the  four-dimensional planes, associated
with the kernels of the linear maps, at infinity   are
$$
\eqalign{
n^{(\mu)}&=g_\infty^{i\lambda, j\rho}
 \partial_{i\lambda} \tau^\mu(p,a) \partial_{j\rho}
\cr
m^{(\nu)}&=g_\infty^{i\lambda, j\rho} 
\partial_{i\lambda} \tau^\nu(q,b)
\partial_{j\rho}\ .}
\eqn\normalv
$$
The angles amongst the normal vectors $n^{(\mu)}$ and $m^{(\nu)}$ are
given by the \lq\lq angles matrix"
$$
\cos(\theta^{\mu\nu})={g_\infty^{i\lambda, j\rho}
  n^{(\mu)}_{i\lambda}
m^{(\nu)}_{j\rho}\over |n^{(\mu)}| |m^{(\nu)}|}
\eqn\anglesf
$$
where
$$
|n^{(\mu)}|^2=g_\infty^{i\lambda, j\rho}  n^{(\mu)}_{i\lambda}
n^{(\mu)}_{j\rho}=\sqrt{g_\infty^{i\lambda, j\rho} p_{i\lambda} p_{j\rho}}
\eqn\nnone
$$
is the length of the vector $n^{(\mu)}$ as
 measured by the asymptotic metric and
similarly for $|m^{(\nu)}|$. To find the 
angles in terms of the parameters
$(\{p_i\}, \{q_j\})$ of the solution, we
 simply substitute the expressions of the
normal vectors \normalv\ into \anglesf.
Thus we find
$$
\cos(\theta^{\mu\nu})={g_\infty^{i\lambda, j\rho} 
p_{i\lambda} q_{j\rho} \delta^{\mu\nu}
+ (g_\infty)^{i\lambda, j\sigma} 
(J^c)^\rho{}_\lambda p_{i\rho} q_{j\sigma} I_c{}^{\mu\nu}
\over \sqrt{g_\infty^{i\alpha,
j\beta} p_{i\alpha} p_{j\beta}}
\sqrt{g_\infty^{k\alpha', \ell\beta'} q_{k\alpha'} q_{\ell\beta'}}}
\eqn\fourangless
$$
>From this, we conclude that, for generic
 choices of the rotational
parameters $\{p,q\}$, there are {\sl four} independent
angles between every pair of intersecting NS-5-branes 
 in the natural coordinates that
we have chosen to express the brane solution. It turns out
that these angles are $Sp(2)$ angles. The most convenient way
to establish this is to use an alternative construction for the
solutions \ans\ which will be presented in [\gpat] and so we shall
not pursue this point further here.

We proceed to find the number of independent
 angles between a {\sl given} pair
of five branes. We shall take them to be
the {\sl minimal} number of independent
eigenvalues of the angles matrix 
 $A=\{\cos(\theta^{\mu\nu})\}$ over the different
parameterizations of the four-dimensional planes \poles.  
To diagonalize the angles matrix in our case, we
write its elements as
$$
{\rm cos}(\theta^{\mu\nu})= 
a\delta^{\mu\nu}+b^c I_c{}^{\mu\nu}
\eqn\nntwo
$$
where $\{a,b^c; c=1,2,3\}$ can
 be easily computed from \fourangless. Then
we choose a complex basis
with respect to the complex structure
$$
K= {b^c\over |b|} I_c\ ,
\eqn\nnthree
$$
where $|b|^2=\delta_{ac} b^a b^c$.
Since $\delta_{\mu\nu}$ is
 hermitian with respect to $K$, we have
$$
{\rm cos}(\theta^{\alpha\bar\beta})= 
(a+i|b|)\delta^{\alpha\bar\beta}\ ,
\eqn\nnfour
$$
where $a, |b|$ depend on the form of the asymptotic metric
and the paramerization of the four-dimensional planes \poles.
For generic asymptotic metric it seems that there is no
relation between  $a$ and $|b|$ and hence 
there are two independent
angles. For Euclidean asymptotic metric,
 if we place the first four-dimensional
plane along the first four coordinates 
of $\bE^8$ and parameterize the second
one using four rotational parameters, then 
the angles matrix has one independent
eigenvalue.

We remark that the above complex basis that 
diagonalizes the angles matrix depends on the pair
of branes involved in the
 intersection. There is no choice of basis that
diagonalizes simultaneously 
the angles matrix of all pairs. 
Hence there are in general 
four angles parameterizing 
the relative position of
the branes involved in the configuration.

\chapter{Other IIA solutions}

The solution found in  section two 
can be superposed with
a fundamental string and a ten-dimensional
 pp-wave. The resulting solution
is
$$
\eqalign{
ds^2&= F^{-1}\big (dudv+(K-1) du^2)+ ds^2_{(8)}
\cr
e^{2\phi}&= F^{-1}  (g_{(8)})^{1\over4}
\cr
H&=du\wedge dv\wedge dF^{-1}+H_{(8)}}
\eqn\wavestring
$$
where $F,K$ are harmonic-like functions satisfying,
$$
\partial_{i\mu}\big(  (g_{(8)})^{1\over4} 
g^{i\mu,j\nu} \partial_{j\nu} F\big)=0
\eqn\harm
$$ 
and similarly for $K$, $u,v$ are light-cone coordinates.
Using   \keyrel, we can rewrite \harm\ as
$$
g^{i\mu,j\nu} \partial_{i\mu} \partial_{j\nu} F=0\ .
\eqn\hamham
$$
A class of solutions of this equation is given by
$$
F=1+\sum_\tau {\mu(\tau)\over |\tau(x)|^2}\ . 
\eqn\nnsix
$$
We note that  the parameters $\{p,a\}$ of the linear maps $\tau$ in the 
above sum are not necessarily
 those that have appeared in the sum for the  metric \fhkt.

Another possibility is to superpose D-4-branes 
at $Sp(2)$-angles intersecting
on a 0-brane. In addition, we can place a 
D-0-brane on this configuration. The
resulting solution is 
$$
\eqalign{
ds^2&= F^{{1\over2}}\big( g_{(8)}^{-{1\over8}}[-F^{-1}dt^2+ds^2_{(8)}]+
g_{(8)}^{{1\over8}} dz^2 \big)
\cr
e^{2\phi}&=  F^{{3\over2}}  (g_{(8)})^{-{1\over8}}
\cr
G_2&= dt\wedge dF^{-1}
\cr
G_4&= H_{(8)}\wedge dz\ ,}
\eqn\zerostring
$$ 
where $F$ is a harmonic-like function,  as in \hamham, 
associated with the
D-0-brane, and $G_2$ and $G_4$  are the IIA Ramond$\otimes$Ramond 
two-and four-form field strengths, respectively. All the 
above solutions preserve at
least
 $3/32$ of
spacetime supersymmetry

\chapter{IIB 5-branes at Angles}

The IIA supergravity solution described in section two
 is also a solution of IIB supergravity.
Alternatively one can use T-duality along the
string direction to transform the IIA solution to a IIB one.
The field equations of IIB supergravity 
are invariant under $SL(2, \bR)$.
Under the action of $SL(2,\bR)$, the metric 
in the Einstein frame remains
invariant, the two three-form field strengths 
$H^1, H^2$ of IIB supergravity
transform as doublets, and 
$$
\tau=\ell+i e^{-\phi_B}
\eqn\llone
$$
transforms by fractional linear transformations,
where $\ell$ is the axion and  $\phi_B$ is the IIB dilaton (see e.g. [\ptg]).
To proceed, we choose $H^1=H$ and $H^2=H'$ the 
three-form field strengths associated with
NS-5-brane and the D-5-brane, respectively.
This symmetry can be used to construct new
 solutions from the one in \ans.
In particular, it is known that under S-duality 
(which is an element of $SL(2, \bR)$),
 the NS-5-branes transform to D-5-branes.
 Therefore it is expected that the solution
 \ans\ as a solution of IIB
supergravity transformed under S-duality will 
lead to a new solution of IIB with the
interpretation of D-5-branes intersecting at 
$Sp(2)$-angles on a string. To perform the
S-duality transformation, we first write the 
metric  \ans\ in the Einstein frame using
$$
ds^2_E=e^{-{1\over2}\phi_B}ds^2_B\ .
\eqn\llfour
$$
Then we apply S-duality to find
$$
\eqalign{
ds^2_E&= g_{(8)}^{{1\over16}} \big(ds^2(\bE^{(1,1)}) +ds^2_{(8)}\big)
\cr
e^{8\phi_B}&=g_{(8)}^{-1}
\cr
H'&=H_{(8)}\ .}
\eqn\llfive
$$

More solutions in IIB can be found by 
T-dualizing the solutions \wavestring\ and
\zerostring\ of IIA theory. In the former case, 
T-duality along the string direction will lead to a solution
in IIB with the same interpretation as that 
in IIA. In the latter case, T-duality along $z$ will lead
to a configuration with the 
interpretation of intersecting
D-5-branes at $Sp(2)$ angles on 
a string superposed with a D-string at the
intersection.

\chapter{M-theory}

The IIA supergravity solution described in  section 
three can be easily  lifted
to M-theory. Let $z$ be the 
eleventh coordinate. The relevant Kaluza-Klein 
ansatz\foot{We have not included in this ansatz 
the Kaluza-Klein vector and the IIA four-form 
field strength because they vanish for our
ten-dimensional solutions.} for the reduction 
from eleven dimensions to ten is
$$
\eqalign{
ds^2_{(11)}&= e^{{4\over3}\phi} dz^2+ e^{-{2\over3}\phi} ds^2_{(10)}
\cr
G_4&=H\wedge dz\ ,}
\eqn\hhone
$$
where $\phi$ is the ten-dimensional dilaton, 
the ten-dimensional metric $ds^2_{(10)}$ is in the
string frame, $G_4$ is the 4-form field
 strength of eleven-dimensional
 supergravity and $H$
is the ten-dimensional  NS$\otimes$NS three-form field strength. Lifting the 
solution \ans\ of IIA theory to
eleven-dimensions, we find the M-theory solution 
$$
\eqalign{
ds^2&= (g_{(8)})^{{1\over6}} dz^2+ 
 (g_{(8)})^{-{1\over12}}\big(ds^2(\bE^{(1,1)})+ ds^2_{(8)}\big)
\cr
G_4&=H_{(8)}\wedge dz}
\eqn\msol
$$
This solution has the interpretation of 
intersecting M-5-branes at
$Sp(2)$-angles on a string separated along the direction $z$. We remark 
though that the
solution is not localized in the $z$ direction.

The above solution can be 
superposed with a membrane without breaking any more 
supersymmetry. The membrane
directions are those of the string 
intersection and that of $z$. 
The resulting solution  is
$$
\eqalign{
ds^2&=F^{-{2\over3}}  (g_{(8)})^{{1\over6}} dz^2+ 
(g_{(8)})^{-{1\over12}}\big[ F^{-{2\over3}} ds^2(\bE^{(1,1)})+
F^{1\over3} ds^2_{(8)}\big]
\cr
G_4&=H_{(8)}\wedge dz+ {\rm Vol}(\bE^{(1,1)})\wedge dF^{-1}\wedge dz}
\eqn\msolb
$$
where  $F$ is a harmonic-like function as in \hamham\ associated
with the membrane.

Apart from these M-brane configurations we 
can allow a pp-wave to propagate along the
string direction. The resulting M-theory configuration is
$$
\eqalign{
ds^2&=F^{-{2\over3}} (g_{(8)})^{{1\over6}} dz^2+
\cr & 
(g_{(8)})^{-{1\over12}}\big( F^{-{2\over3}} (du dv+(K-1) du^2)+
F^{1\over3} ds^2_{(8)}\big)
\cr
G_4&=H_{(8)}\wedge dz+ {\rm Vol}(\bE^{(1,1)})\wedge dF^{-1}\wedge dz}
\eqn\msolc
$$
where $F,K$ are harmonic-like functions as in \hamham.
This solution includes  all previous ones and
preserves $3/32$ of the supersymmetry. For example if
$F$ and $K$ are one, then we recover 
the solution \msol. Setting $K=1$,
we recover the solution \msolb.
 We can also set $F=1$ in which case we find
a new M-theory solution which has the 
interpretation of M-5-branes intersecting on
a string at $Sp(2)$-angles and a wave
 propagating along the string. 
Reduction of this solution along the 
pp-wave direction gives the solution \zerostring\
of the IIA theory. We can also reduce \msolc\ along the same
direction yielding a new IIA solution. We expect that 
 our solutions will  receive corrections due to the 
anomaly terms induced by the M-5-brane to the D=11
supergravity action as those in [\marica].

\chapter{Conclusions}

We have constructed new solutions
 with the interpretation of intersecting
IIA NS-5-branes at $Sp(2)$-angles on
 a string preserving at least $3/32$ 
of  supersymmetry. We have shown that there are four
independent angles between every
 pair of intersecting NS-5-branes,
respectively.
We have described  the superposition of the
intersecting NS-5-brane solutions with a fundamental string and a
pp-wave. We have also investigated 
the T-duals of these solutions as well as their
M-theory interpretation. 

The intersecting IIA NS-5-brane
solutions that we have found 
 are also solutions of the
heterotic and type I strings. 
It would be of interest to 
investigate further our brane
solutions  in the context 
of the heterotic string using as 
 Yang-Mills fields the instantons of [\kent] (see [\lambert]). It is expected
that consideration of the 
cancellation of chiral
 anomalies of the heterotic string
will modify our solutions.
 A related problem is the 
investigation
of heterotic sigma models 
with bosonic couplings given by the 
geometries found in section 
two and with Yang-Mills couplings provided
by the instantons of [\kent]. 
These sigma models admit a (4,0)-supersymmetric
extension. Therefore they
 are expected to be  
ultraviolet finite [\howea,  \howeb].  However
due to the presence of 
sigma model anomalies, their couplings may receive
$\alpha'$ corrections.

\vskip 0.2cm
\noindent{\bf Acknowledgments:}  We would like to thank G.W. Gibbons
and P.K. Townsend for helpful discussions. Part of this work
was  done during the visit of one of us,
G.P., at the
Institute for Theoretical Physics of the University
 of California, Santa Barbara. G.P
thanks  the organizers of {\sl Dualities in String Theory}
programme,  in particular M. Douglas, for an
 invitation to visit the Institute. A.T.
thanks PPARC for a  studentship. G.P. is 
supported by a University Research Fellowship
from the Royal Society. This research was supported
in part by the National Science Foundation under Grant
No. PHY94-07194.

\refout
\bye